\documentclass[9pt,twocolumn,twoside,lineno]{pnas-new}
\usepackage{xr}
\templatetype{pnasresearcharticle} 

\title{Origin of filaments in finite-time in Newtonian and non-Newtonian thin-films}

\author[a]{Saksham Sharma}
\author[a]{D. Ian Wilson} 

\affil[a]{Department of Chemical Engineering and Biotechnology, University of Cambridge,
Philippa Fawcett Drive, Cambridge CB3 0AS, UK}

\leadauthor{Sharma \& Wilson} 

\correspondingauthor{\textsuperscript{2}To whom correspondence should be addressed. E-mail: ss2531@cam.ac.uk}


\begin{abstract}
The sticky fluids found in pitcher plant leaf vessels can leave fractal-like filaments behind when dewetting from a substrate. To understand the origin of these filaments, we investigate the dynamics of a retreating thin-film of aqueous polyethylene oxide (PEO) solutions which partially wet polydimethyl siloxane (PDMS) substrates. Under certain conditions the retreating film generates regularly-spaced liquid filaments. The early-stage thin-film dynamics of dewetting are investigated to identify a theoretical criterion for liquid filament formation. Starting with a linear stability analysis of a Newtonian or simple non-Newtonian (power-law) thin-film, a critical film thickness is identified which depends on the Hamaker constant for the fluid-substrate pair and the surface tension of the fluid. When the measured film thickness is smaller than this value, the film is unstable and forms filaments as a result of van der Waals forces dominating its behaviour. This critical film-height is compared with experimental measurements of film thickness obtained for receding films of Newtonian (glycerol-water mixtures) and non-Newtonian (PEO) solutions generated on substrates inclined at angles $0^{\circ}$, $30^{\circ}$, and $60^{\circ}$ to the vertical. The observations of filament and its absence show good agreement with the theory. Further analysis of the former case, involving a stability analysis of the contact line, yields a prediction of the spacing (wavelength) $\hat{\lambda_{f}}$ between filaments as $\hat{\lambda}\textsubscript{f}\hat{\eta}/\hat{\gamma} \propto Ca$, where $\hat{Ca}$ is the capillary number for contact line motion: our experiments yield $\hat{\lambda}\textsubscript{f}\hat{\eta}/\hat{\gamma} \propto Ca^{1.08}$ and earlier studies in the literature reported $\hat{\lambda}\textsubscript{f}\hat{\eta}/\hat{\gamma} \propto Ca^{0.945}$ . The evolution of the thin-film shape is modelled numerically to show that the formation of filaments arises because the thin-film equation features a singular solution after a finite-time, hence termed a ``finite-time singularity".  
\end{abstract}

\dates{This manuscript was compiled on \today}
\doi{\url{www.pnas.org/cgi/doi/10.1073/pnas.XXXXXXXXXX}}

\begin{document}

\maketitle
\thispagestyle{firststyle}
\ifthenelse{\boolean{shortarticle}}{\ifthenelse{\boolean{singlecolumn}}{\abscontentformatted}{\abscontent}}{}

\begin{figure*}
  \centering
  \includegraphics[width=0.9\textwidth]{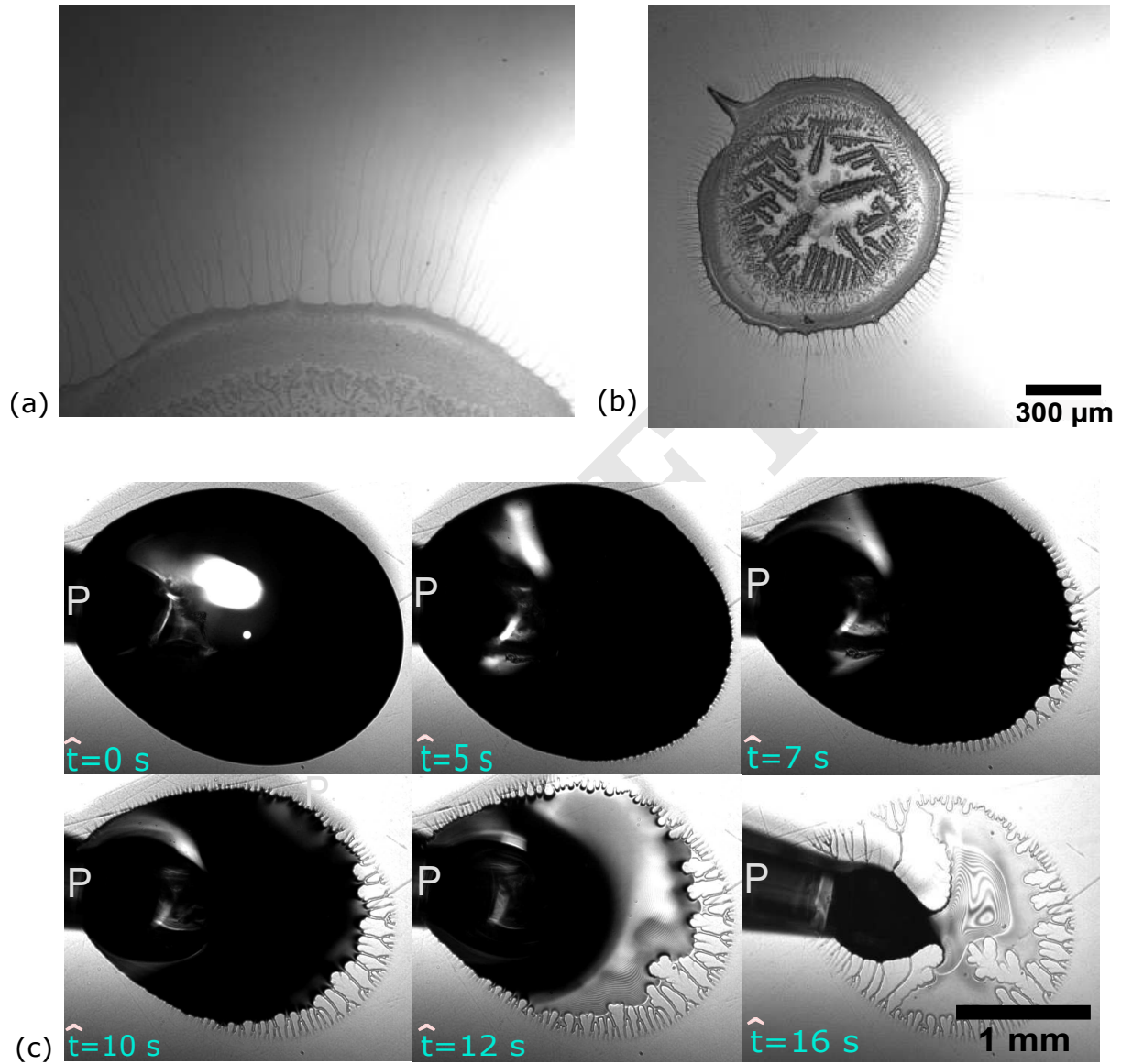}
 \caption{Filaments formed by overnight evaporation of ground pitcher \textit{N. Rafflesiana} fluid  on (a) polystyrene and (b) borosilicate glass surfaces. (c) Forced shrinkage of a sessile drop of \textit{N. {R}afflesiana} on borosilicate glass, caused by withdrawal of liquid via pipette labelled P. Evaporation in (a) and (b) results in concentration of dissolved species in the drop, slowing evaporation and triggering the transition from a receding film to a pinned state and ‘coffee ring effect’ features. Liquid removal in (c) is faster and there is little evaporation: filaments are observed once the contact line starts to recede.} 
  \label{fig:fig0}
\end{figure*}

Receding thin films arise in a range of fields and are important in coating, printing and drying applications, where the stability of the thin film near the moving contact line will determine the uniformity of the product. Our interest in this topic arises from observation of residual filaments left by the evaporation of sessile droplets of the sticky digestive fluid secreted by \textit{N. Rafflesiana} pitcher plants. Fig. \ref{fig:fig0} shows that these filaments are formed in the early stages of evaporation, where the shrinking drop forms a receding thin film at the contact line, and can be generated artificially by sucking liquid from the drop (Fig. \ref{fig:fig0}(c)). The filaments exhibit regular spacing on relatively smooth surfaces, indicating that the filaments arise from an instability in the thin film rather than contact line pinning. \newline

Pitcher plant fluids are non-Newtonian aqueous solutions containing long-chain polysaccharides \citep{adlassnig2010deadly}. Deblais et al. \citep{deblais} observed similar filaments with thin films of viscous Newtonian and non-Newtonian liquids (glycerine and synthetic polymer solutions, respectively) generated by a blade arrangement which allowed the initial height and contact line velocity $\hat{U}$ to be controlled independently. As the contact line velocity $\hat{U}$ decreased, uniform films with a straight contact line were replaced by ones with regularly spaced cusps and rivulets: for Newtonian liquids the rivulets were unstable and gave rise to droplets whereas the higher extensional viscosity of the polymer solutions stabilised the filaments and gave patterns analogous to those in Fig. \ref{fig:fig0}(c)). They reported that the threshold of cusp (and filament) formation corresponded to a critical value of the capillary number, $Ca=\hat{\eta} \hat{U} /  \hat{\gamma}$, where $\hat{\eta}$ is the apparent viscosity and $\hat{\gamma}$ the surface tension, but did not provide a theoretical treatment. \newline

In the present work, we provide a theoretical explanation for the observed filament formation. It draws on recent experimental work by Xue \& Stone on a liquid film draining down a glass slide under the action of gravity. The thin-film non-linear PDE used to model the film considered three forces: viscous resistance, gravity, and surface tension. A major assumption there was the consideration of perfectly wetting (zero degree contact angle) fluids, arising from the difficulty of incorporating partial wetting in the model, as remarked by co-author Stone when presenting this work at the \href{https://youtu.be/s4Sqml7Kcec?t=3230}{\textcolor{blue}{GKB 100 symposium}}~\citep{ytstone}. Partial wetting is included in the present study, following the work by Witelski and coworkers on the stability analysis and evolution of thin-films (\cite{witelski2000}, \cite{witelski2001}, \cite{witelski2008} and \cite{witelski2020}) by employing expressions for the van der Waals forces' dependency on the film thickness. We demonstrate that the formation of cusps and filaments observed during the dewetting of pitcher plant fluids has its roots in the inherent instability of the thin-film. Strictly speaking, this instability gives rise to a `finite-time singularity'. \newline

The article is organised as follows. The hydrodynamic equations for the thin-film are presented in Sec. \ref{sec:theory} and stability analysis is then performed to find the critical criterion for filament formation (Sec. \ref{sec:stability}). The rationale behind the finite-time singularity feature of the thin-film equation is discussed in Sec. \ref{sec:finitetime} along with some numerical investigations. The criterion is compared with experimental results in Sec. \ref{sec:results}. A scaling law for the spacing between filaments is derived and compared with experimental data in Sec. \ref{sec:mcl}. The key findings and potential further directions for this work are discussed in Sec. \ref{sec:discussion}. \newline

\section{Theory}\label{sec:theory}
\begin{figure*}
  \centering \includegraphics[width=0.9\textwidth]{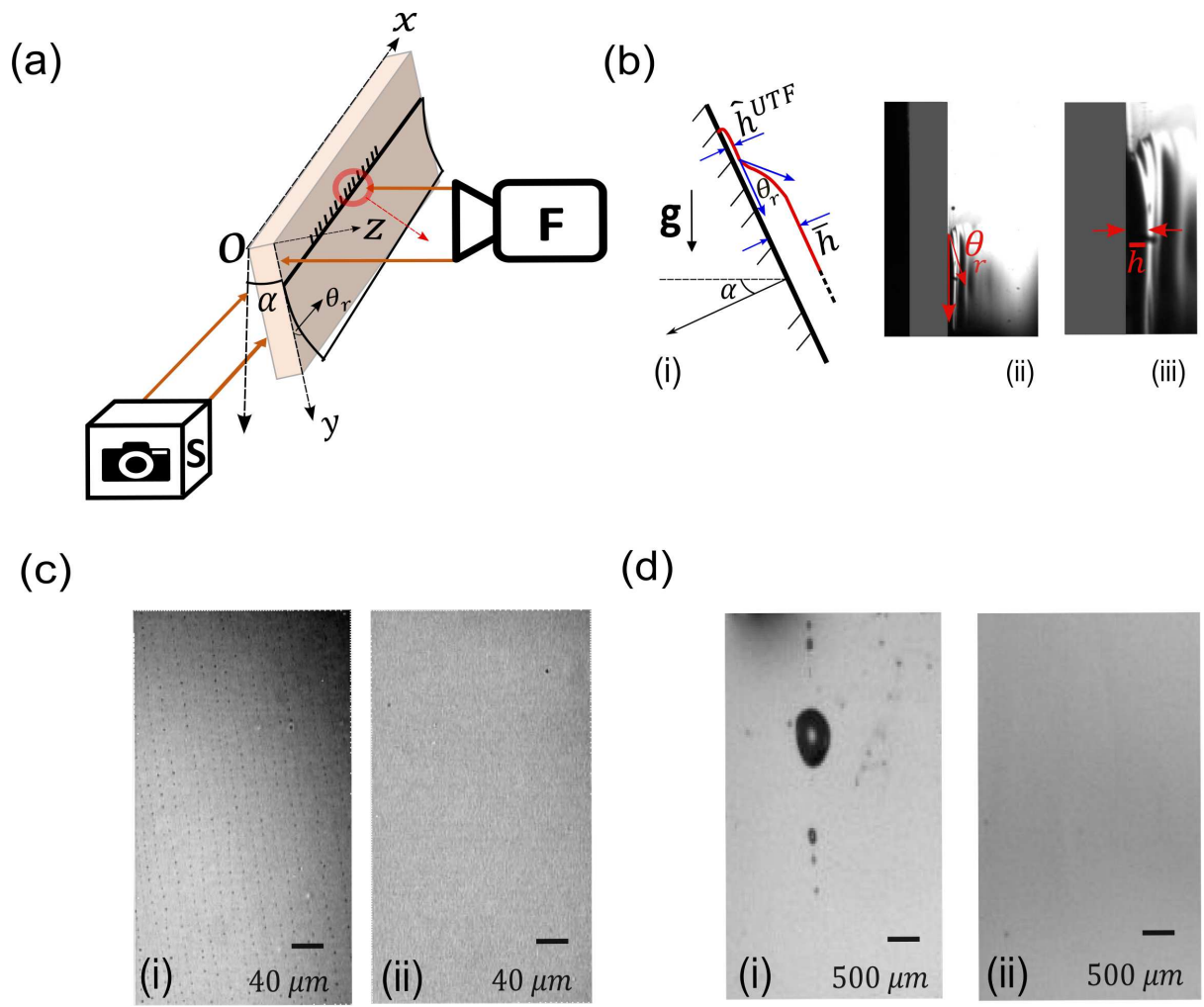}
 \caption{(a) Schematic of experimental setup showing the coated glass microscope slide on which a thin-film dewets as it drains downwards. Cameras $F$ (xiQ USB3.0) and $S$ (acA1300-200um) record the dewetting process from the front and side, respectively. Cartesian co-ordinates indicated by axes. (b) Side view of the receding film - (i) schematic and (ii) camera image - with receding contact angle $\theta_{r}$, film heights $\hat{h}^{UTF}$ and $\bar{h}$ (scale: width of glass slide is 1 mm shown by the shaded grey region). (c) Microscope images of the substrate (i) with and (ii) without filaments during dewetting of PEO solution. (d) Front (camera $F$) view of substrate (i) with and (ii) without droplet formation during dewetting of aqueous glycerol solution.}
  \label{fig:fig1}
\end{figure*}

Consider a thin liquid film of local thickness $\hat{h}(\hat{x},\hat{y},\hat{t})$ on the plane $(Oxy)$ as shown in Fig. \ref{fig:fig1}(a)), with height $\hat{h}$ pointing towards $Oz$. $Ox$ points horizontally along the liquid-substrate-air contact line and $Oy$ points directly down the slope. A standard thin-film equation for a non-Newtonian liquid \cite{nasehi2018evolution} exhibiting power-law behaviour with exponent $n$ is
\begin{equation} \label{eq:thinfilm}
    3\hat{\eta} \frac{\partial \hat{h}}{\partial \hat{t}} = \left ( \frac{\partial}{\partial \hat{x}} \left[\hat{m}(\hat{h})\frac{\partial}{\partial \hat{x}} (\hat{p})\right] - 3\hat{\rho} g'\hat{h}^{2}\frac{\partial \hat{h}}{\partial \hat{y}} \right) ^{1/n}  
\end{equation}
Here $\hat{\eta}$ is short-form for apparent viscosity $\hat{\eta}(\dot{\gamma})$, where $\dot{\gamma}$ is the shear-rate of the thin-film defined at its free surface and given as $\hat{U}/\bar{h}$ (see Fig. 
 \ref{fig:fig1}(b.ii)), $\hat{\rho}$ the density, $\hat{m}(\hat{h})$ the mobility coefficient, $\hat{p}$ is the hydrodynamic pressure $\hat{p}(\hat{x},\hat{t})$ of the thin-film, and $g'=g \cos \alpha$; hat denotes that the term is dimensional. The boundary condition at the film-substrate interface (on surface $Oxy$) gives $\hat{m}(\hat{h})=\hat{h}^{k}$ where $k \in [1,3]$. The pressure $\hat{p}(\hat{x},\hat{t})$ in the thin-film \cite[see Eq. (1.3)]{witelski2008} is given by 
\begin{equation} \label{eq:pressure}
    \hat{p} = \hat{\Pi}(\hat{h})-\hat{\gamma} \frac{\partial^{2}\hat{h}}{\partial \hat{x}^{2}}  
\end{equation}
where $\hat{\Pi}(\hat{h})$ is the disjoining pressure which accounts for the van der Waals interactions between the thin-film and the substrate. The second term on the RHS accounts for the effect of surface tension $\hat{\gamma}$ and curvature of the thin-film. The disjoining pressure $\hat{\Pi}(\hat{h})$ can be written in the form 
\begin{equation} \label{eq:pressure2}
    \hat{\Pi}(\hat{h})=\frac{\hat{A}}{\hat{h}^{3}}\left[1-\frac{\hat{h}_{UTF}}{\hat{h}} \right]  
\end{equation}
where van der Waals forces are characterised by the Hamaker constant $\hat{A}$ ($\hat{A}>0$ means the interaction is hydrophobic and $\hat{A}<0$ hydrophilic) and $\hat{h}_{UTF}$ is the height of the adsorbed precursor film in Fig.\ref{fig:fig1}(b)) \citep{witelski2008}. There are two timescales in \eqref{eq:thinfilm}: i) $\hat{T_{x}}$, when the pressure term (first term on the RHS) is dominant, and ii) $\hat{T_{y}}$, when the gravity term (second term on the RHS) dominates. A scaling analysis gives these timescales as 
\begin{equation} \label{eq:timescales}
    \hat{T_{x}}=\frac{3\hat{\eta} \hat{L_{x}}^{4}}{\bar{h}^{3}\hat{\gamma}}; \quad \hat{T_{y}}=\frac{\hat{\eta} \hat{L_{y}}}{\hat{\rho} g' \bar{h}^{2}}
\end{equation}
with $\hat{L_{x}}=\bar{h}\sqrt{\hat{\gamma} \hat{h}_{UTF}^{2}/\hat{A}}$ (\citep[see][p.~016301-2]{witelski2008}), and length scale $\hat{L_{y}}$ the length of the glass substrate in the $Oy$ direction (25 mm in our tests). With $\bar{h}=O(10^{-4})$ m, $\hat{A}=O(10^{-17})$ J, $\hat{h}_{UTF}=O(10^{-9})$ m and $\hat{\gamma}=O(10^{-2})$ N/m, this gives $\hat{T_{x}}=O(10^{-8})$ s and $\hat{T}_{y}=O(1)$ 
so the first term on the RHS in \eqref{eq:thinfilm} is expected to be dominant. \newline

This means that early-stage dynamics of the thin-film, when surface tension and van der Waals interactions dominate, are characterised by a time scale of $10^{-8}$ s, compared to the intermediate stage where gravity is important. Similar arguments about time scales have been reported previously \citep[p.~016301-2]{witelski2008}. It means that, as soon as the thin film is deposited on the substrate (or, in these experiments, the substrate is withdrawn from the liquid pool), an interplay between surface tension and van der Waals interaction forces begins. \newline

The experiments reported here employed Newtonian solutions (mixtures of glycerol and water) of different viscosity as well as non-Newtonian ones (mixtures of polyethylene oxide, PEO, in water). Scaled viscosity is used to label the liquids when presenting results, given by $\hat{\eta}$/$(\hat{\rho} \hat{\gamma}^{3}/g')^{1/4}$. The PEO solutions exhibit shear-thinning, which can be described by the Cross model (see Supplementary Material). The shear rate in the experimental films lie in the power law regime for this model (see Supp. Fig. S1), which is why this model is used in \eqref{eq:thinfilm}. \newline

Hence, in the next section, we focus on the early-stage dynamics of the thin-film by ignoring the gravity term in \eqref{eq:thinfilm}, \textit{viz}. 
\begin{equation} \label{eq:maineqn}
    3\hat{\eta} \frac{\partial \hat{h}}{\partial \hat{t}} = \frac{\partial}{\partial \hat{x}} \left[\hat{h}^{k}\frac{\partial}{\partial \hat{x}} \left(\frac{\hat{A}}{\hat{h}^{3}}\left[1-\frac{\hat{h}_{UTF}}{\hat{h}} \right]-\hat{\gamma} \frac{\partial^{2}\hat{h}}{\partial \hat{x}^{2}}  \right)\right]^{1/n} .
\end{equation}
\noindent
\eqref{eq:maineqn} is non-dimensionalised by introducing scales
\begin{equation} \label{eq:scales}
    \hat{h}=\bar{h}h, \quad \hat{x}=\hat{L_{x}}x, \quad \hat{t}=\hat{T_{x}}t, \quad \hat{h}_{UTF}=\bar{h}\zeta
\end{equation}
\noindent
where the variables without hats are dimensionless and $\zeta=\hat{h}_{UTF}/\bar{h}<1$. This yields
\begin{equation} \label{eq:dimeqn}
  \frac{\partial h}{\partial t} = \frac{\partial}{\partial x} \left[h^{k}\frac{\partial}{\partial x} \left(\Gamma(h)-\frac{\partial^{2}h}{\partial x^{2}} \right)\right]^{1/n}   
\end{equation}
where $\Gamma(h)$ is the dimensionless form of $\hat{\Pi}(h)$ and can be written as
\begin{equation}
    \Gamma (h)=\frac{\zeta^{2}}{h^{3}}\left[1-\frac{\zeta}{h}\right]
\end{equation}
and the dimensionless pressure term is
\begin{equation}
    p = \Gamma(h)-\frac{\partial^{2} h}{\partial x^{2}}
\end{equation} 
which will be used in the next section to simplify the analysis of this PDE.

\section{Stability analysis of the steady state solution of the thin-film PDE}\label{sec:stability}
We investigate the tendency of a thin-film to maintain (or lose its) stability at an early stage. Since the formation of filaments happens almost instantaneously, the phenomenon is primarily linked to the interplay of surface tension and van der Waals interaction forces. A linear stability analysis is performed with respect to arbitrary infinitesimal perturbations. We start by finding a Lyapunov function for \eqref{eq:dimeqn} - in effect, modelling it as a dynamic system - because the existence of such a function provides an indication of the nature of stability of the system \citep{strogatz}. Consider the following integral 
\begin{equation} \label{eq:lyapunov}
  I[h]=  \int_{0}^{1} \left( \frac{1}{2} \left|\frac{\partial h}{\partial x}\right|^{2}- \varsigma(h)\right) dx 
 \end{equation} 
where $\partial \varsigma(h)/\partial h=\Gamma(h)$. The reason behind choosing this integral is that its first derivative w.r.t $h$, has the property
\begin{equation} \label{eq:firstvar}
  \frac{\partial I}{\partial h}= \frac{\partial}{\partial h} \left(\frac{1}{2} \left|\frac{\partial h}{\partial x}\right|^{2}- \varsigma(h)\right) =  \frac{\partial^{2}h}{\partial x^{2}} - \Gamma(h) = -p.
\end{equation}
Such an integral is referred to as an energy functional of the system. The rate of energy dissipation in a control volume $dV$ for this energy functional is given by 
\begin{equation} \label{eq:localenergydiss}
  \frac{\partial I}{\partial t}= \frac{\partial I}{\partial h} \frac{\partial h}{\partial t} = -p \ \nabla.\left(h^{3}\nabla p \right)
\end{equation}
 where the derivative is written in terms of operator $\nabla$ to simplify further analysis.
 Total energy dissipation over volume $V$ is
 \begin{equation} \label{eq:totenergydiss1}
  \frac{d I}{d t}= \int_{V}  -p \ \nabla.\left(h^{3}\nabla p \right) dV
\end{equation}
Integrating by parts gives 
 \begin{equation} \label{eq:totenergydiss2}
  \frac{d I}{d t}= -p\int_{V} \nabla.\left(h^{3}\nabla p \right) dV - \int_{V} \nabla p \int \nabla.(h^{3}\nabla p) dV
  \end{equation}
 and applying the divergence theorem gives
 \begin{equation} \label{eq:totenergydiss3}
  \frac{d I}{d t}= -p\int_{S} h^{3}(\nabla p.\mathbf{n}) dS - \int_{S} \ h^{3} (|\nabla p|^{2}.\mathbf{n}) \ dS.
\end{equation}
The first term on the RHS is zero because of the boundary condition which assumes no flux of the liquid normal to the boundary, \textit{i.e.} $\nabla p.\mathbf{n}=0$ on surface $dS$ (the liquid/vapour interface). As a result, the expression for net energy dissipation is
\begin{equation} \label{eq:totenergydiss4}
  \frac{d I}{d t}= -\int_{S} \ h^{3} |\nabla p|^{2}\ dS
\end{equation}
which is always $\leq 0$ because $h \geq 0$. The non-increasing nature of Lyapunov function $I[h]$ confirms the direction of stability of the steady state solution of \eqref{eq:dimeqn}, which is towards the higher value of $h$. This approach allows one to find all possible time-independent solutions of \eqref{eq:dimeqn} which obey \eqref{eq:totenergydiss4}, and these solutions are: (\textit{i}) $h=0$ and (\textit{ii}) $\nabla p  =0$. The former corresponds to the absence of any fluid on the surface and the latter means that a uniform pressure field exists for the steady-state solution. The steady-state solution with uniform pressure $\Bar{p}$ is selected to proceed further, such that
\begin{equation} \label{eq:uniform}
    \Gamma(h)-\frac{\partial^{2}h}{\partial x^{2}} = \Bar{p} ,
\end{equation}
which has a uniform solution given by $h=\Bar{h}$ and $\Bar{p}=\Gamma(\Bar{h})= \frac{\zeta^{2}}{\Bar{h}^{3}}\left[1-\frac{\zeta}{\Bar{h}}\right] > 0$. It will be demonstrated in the next Section that such a solution can evolve to approach the first solution, $h=0$, for certain parametric conditions. For now, we are interested in finding families of solutions (apart from $h=\Bar{h}$) of \eqref{eq:dimeqn} which can also exist if the original equation is perturbed. To do this, we expand the pressure and the film thickness as 
\begin{subequations}
\begin{gather}
    \Bar{p}=\Bar{p}_{c}+\delta \\ 
    h(x)=\Bar{h}+\epsilon h_{1}(x)+\epsilon^{2}h_{2}(x)+...
\end{gather}
\end{subequations}
\noindent
where $\epsilon, \delta << 1$ are small perturbation parameters. At $O(\epsilon)$, \eqref{eq:uniform} becomes
\begin{equation}\label{eq:order_eps}
    \frac{3 \zeta^{2} h_{1}}{\Bar{h}^{4}} \left[1-\frac{\zeta}{\Bar{h}} \right] + \frac{\partial ^{2} h_{1}}{\partial x^{2}}= 0
\end{equation}
which is a second-order ODE when higher order $\epsilon^{2}$ terms are ignored. The equation is made closed-form by considering it over a periodic domain, $0 \leq x \leq 1$, with Neumann boundary conditions: $h_{x}(0)=h_{x}(1)=0$. Using the periodic boundary condition allows one to explain the formation of a periodic array of filaments in the $Ox$ direction. \eqref{eq:order_eps} can be written as a harmonic oscillator equation of the form
\begin{equation}
    \frac{\partial ^{2} h_{1}}{\partial x^{2}} + \Lambda^{2} h_{1}=0
\end{equation}
where $\Lambda^{2}=3 \zeta^{2} [1-\zeta / \Bar{h}]/ \Bar{h}^{4}$. The eigenvalues of this equation are $\Lambda_{i}=i\pi$  for $i \in \mathbb{Z^{+}}$. Solving for $\Bar{h}$, one obtains the quintic relationship
\begin{equation}  \label{eq:quintic}
    3 \frac{\zeta^{2}}{\Bar{h^{4}}} \left[1-\frac{\zeta}{\Bar{h}}\right]= i^{2}\pi^{2}
\end{equation}
\noindent
with the approximate solution 
\begin{equation}  \label{eq:solution}
    \Bar{h}_{i}= 3^{1/4}\sqrt{\frac{\zeta}{i \pi}}
\end{equation}
\noindent
The solution $\Bar{h}_{1}$ for $i=1$ is the primary bifurcation point which demarcates a transition between stable and unstable solutions: above the primary bifurcation point ($\bar{h}>\bar{h}_{1}$), the solution is stable, and below it ($\bar{h}<\bar{h}_{1}$), it is unstable \cite[see][p.~159]{witelski2000}. Higher order bifurcations for $i \geq 2$ are not considered because they involve only unstable modes \citep[see][Theorem 5]{laugesen}. \newline

To find a dimensional version of this solution, dimensional analysis of \eqref{eq:maineqn} is required, which yields the scaling relation $\bar{h} \backsim \hat{\gamma} \hat{T_{x}}/\hat{\eta}$. Using $\hat{T_{x}}$ from \eqref{eq:timescales} and $\hat{L_{x}}=\bar{h}^{2}\zeta \sqrt{\hat{\gamma}/\hat{A}}$, the critical height of the thin-film is then
\begin{equation} \label{eq:criterion}
\hat{h}_{f}=3^{-5/16}\left(\frac{\zeta}{\pi}\right)^{-1/8}\zeta^{-1}\left(\frac{\hat{\gamma}}{\hat{A}}\right)^{-1/2}
\end{equation}
such that if $\bar{h}<\hat{h}_{f}$, the film is unstable, and if $\bar{h}>\hat{h}_{f}$, it is stable. It should be noted that this theoretical treatment does not predict the shape or behaviour of the unstable film: it provides a criterion to compare the steady state film thickness with a theoretical limit, above which the film is stable and below which it is unstable. This analysis is independent of the rheology of the thin-film and the shear rate associated with the motion. The physical reasoning behind this is that the formation of filaments happens over a timescale significantly faster (theoretically, eight orders of magnitude) than the timescale of receding motion, and thus, the shear rate or rheology of the thin-film does not play any role in the critical film thickness $\hat{h}_{f}$. The reader is referred to the work of Xue and Stone \cite{stone} where the evolution of the shape of the thin-film is considered. \newline

An unstable filament evolves to yield the steady-state solution $h \approx 0$ (as discussed above). The dynamics of this transition are considered in the next Section, where it is shown that the evolution of a thin-film to approach zero film thickness in finite-time is the reason for the origin of the filament.

\section{Finite-time filament formation}\label{sec:finitetime}
We start with a simplified form of \eqref{eq:dimeqn} in this section, where $n=1$, 
to study the bounds or limits of the modified equation which has surface tension and van der Waals force terms. The motivation in the section is to understand the behaviour of the solutions and thereby predict the conditions necessary for the solution to remain bounded or to blowup. A general form of  \eqref{eq:dimeqn} with $n=1$  (see \citep{hocherman}) is
\begin{equation} \label{eq:generalform}
    \frac{\partial h}{\partial t}=-\frac{\partial}{\partial x} \left(M(h) \frac{\partial}{\partial x} \left[-Q(h)+R(h)\frac{\partial^{2} h}{\partial x^{2}}\right]  \right)
\end{equation}
with boundary condition: $h(x+L)=h(x)$. If $M(h)=f(h), R(h)=1$ and $dQ(h)/dh=-g(h)/f(h)$, then the equation becomes 
\begin{equation}  \label{eq:4thordernonlinear}
    h_{t}=-(f(h)h_{xxx})_{x}-(g(h)h_{x})_{x}
\end{equation}
where subscripts denote differentiation. For $f(h)=h^{k}$ ($1\leq k \leq 3$) and $g(h)=3\frac{\zeta^{2}}{h^{4}}(1-\frac{4\zeta}{3 h})$, \eqref{eq:4thordernonlinear} is equivalent to \eqref{eq:dimeqn}. It was proved in \citep{bertozzi} that the solution of \eqref{eq:4thordernonlinear} blows up when 
\begin{equation}
    \lim_{s\rightarrow\infty} \frac{s^{2}f(s)}{g(s)} < \infty
\end{equation}
and when $s^{2}f(s)/g(s)\rightarrow\infty$, as $s \rightarrow \infty$, the classical-solution is uniformly bounded and positive. In the current work, \textit{i.e.}
\eqref{eq:dimeqn} with $n=1$, $s^{2}f(s)/g(s) = s^{6+k}/(3\zeta^{2}(1-4\zeta/3s))$, which grows to $\infty$ as $s\rightarrow \infty$. Hence, the classical-positive solution of \eqref{eq:dimeqn} (for $n=1$) is bounded and finite. Having established the bounds of the equation, we next explore how the solution evolves over time. \newline

\begin{figure}
  \centering
\includegraphics[width=0.85\linewidth]{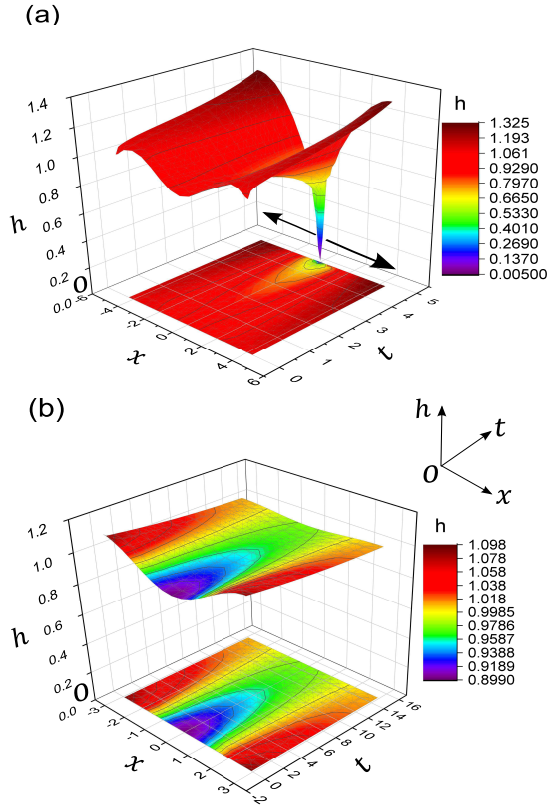}
\caption[Numerical modelling of thin-film evolution]{Numerical solution of  \eqref{eq:maineqn} (for $\hat{h}^{UTF}=0$), over domain $-L_{d}\leq x \leq L_{d}$. (a)  For $L_{d}=4.34$, height $\bar{h}$ approaches zero at time $t_{f} \approx 4.79$ (\textit{finite-time} singularity) after which the dry region grows in the direction of the arrows, and the film splits to form two filaments (not shown). (b) For $L_{d}=2.94$, height $h$ remains stable with time and does not tend towards zero.}
\label{fig:fig2}
\end{figure} 
\eqref{eq:maineqn} states that the height of the film $h$ is determined by two forces: van der Waals interactions and surface tension. For wetting liquids the former is attractive in nature and brings the fluid in contact with the substrate; being inversely proportional to $h$, it increases in magnitude as $h \rightarrow 0$. The latter, on the other hand, tries to minimise the surface energy of the fluid by reducing the area of the liquid-vapour interface; it prefers a non-planar film to a planar one. It is expected that the combination of these will lead to the presence of both features, namely reduced film-height and non-planar film-geometry, in the system. To focus on the highly nonlinear terms (van der Waals and surface tension forces) which grow very fast, we take $\hat{h}^{UTF}=0$ (as the order of $\hat{h}^{UTF}$ is signficantly lower than $h$) in \eqref{eq:maineqn}, so that only the first order term in van der Waals interaction and the surface tension is compared and evaluated in the numerical study. The numerical solution is evaluated using Wolfram Mathematica. A finite domain with width (in the $x$ direction) $2L_{d}$ is considered with the boundary condition $h(-L_{d},t)=h(L_{d},t)$ and  the initial condition $h(x,0)= \Bar{h}- \delta \cos{\frac{\pi x}{L_{d}}}$ such that $\Bar{h}=1$ is the average film height and $\delta=0.1$ is the amplitude of perturbation (0.1 being chosen to ease graphical visualisation). As derived previously in \citep[see Eq. (3.3)]{witelski2020}, there is a critical thickness $\sqrt{L_{d}/\pi}$ (following a similar stability analysis to that in the previous Section) below which the uniform film is unstable and above which the uniform film is stable to small perturbations. \newline 

Fig. \ref{fig:fig2}(a) shows the former case, for $L_{d}=4.34$, > $\pi$, where the film-height $h$ approaches 0 at $t \approx 4.79$ (in finite-time). Near this time, the Mathematica software suspects a singularity to exist, hence it is not possible to find the solution at longer times. Fig. \ref{fig:fig2}(b) shows an example of the latter case, for $L_{d}=2.94$, where no such singularity is found even at larger times; the initial sinusoidal perturbation decays to the average film-height $\Bar{h}=1$, suggesting that the film remains stable.

 \begin{figure*}
  \centering
  \includegraphics[width=1\linewidth]{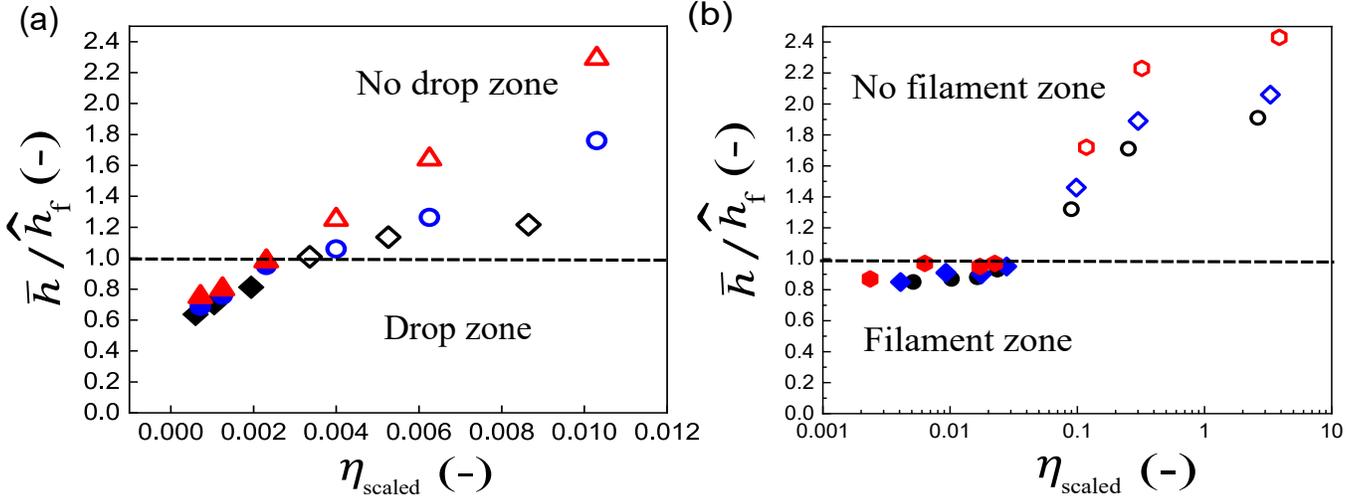}
\caption[Comparison between experiments and results]{Comparison of experimental results with theory: effect of dimensionless viscosity, $\eta_{\text{scaled}}$, on $\bar{h}/\hat{h}_{f}$ for (a) glycerol-water mixtures, and (b) PEO solutions. Solid symbols denote unstable film, forming singularities in the form of drops or filaments; hollow symbols denote a stable film. Symbol colour indicates the angle of plate inclination: \textbf{black} for $\alpha=0^{\circ}$; \textcolor{blue}{\textbf{blue}} for $\alpha=30^{\circ}$; \textcolor{red}{\textbf{red}} for $\alpha=60^{\circ}$. Here, $\bar{h}$ is the experimentally measured film thickness and $\hat{h}_{f}$ is the theoretical critical film thickness. }
\label{fig:hhf_results}
\end{figure*}

\section{Results}\label{sec:results}

In the experiments a receding liquid film is generated on a static coated glass slide by moving the liquid reservoir downwards. The film thickness is measured and the initial behaviour of the receding film is recorded as outlined in Fig. \ref{fig:fig1}(b). Fig. \ref{fig:hhf_results} plots the ratio $\bar{h}/\hat{h}_{f}$, which determines whether the film will remain stable or not, against the dimensionless viscosity, 
$\eta_\text{scaled}$ (scaled with $(\hat{\rho} \hat{\gamma}^{3}/g\sin\alpha)^{1/4})$ for glycerol-water mixtures and aqueous PEO solutions. \newline

Fig. \ref{fig:hhf_results}(a) shows good agreement with the stability criterion for the Newtonian liquids. If $\bar{h}/\hat{h}_{f}<1$, the film becomes unstable because the thinner film gives rise to larger van der Waals forces. A more viscous liquid, \textit{i.e.} larger $\eta_\text{scaled}$, results in a larger film thickness $\bar{h}$ (measured experimentally - see Supplementary Tables S3-S5), while $\hat{h}_{f}$ does not depend on viscosity: less viscous Newtonian films are more likley to be unstable.  An unstable film touches the surface to create a dry spot which grows in the region ~\citep[see Fig.5]{witelski2020} and gives rise to satellite beads. Satellite beads have been reported previously, \textit{e.g.}\citep{limat}, where drops sliding down on a partially wetting substrate emitted smaller drops from their cusped tail. \newline 

For the non-Newtonian case, Fig. \ref{fig:hhf_results}(b), when $\bar{h}/\hat{h}_{f}>1$, the thin-film again remains stable and does not yield a singularity. However, for $\bar{h}/\hat{h}_{f}<1$, the thin-film yields a singular solution and filaments are observed. Experimentally observed zones of filament formation and non-formation are predicted very well by the stability theory. The crossover between the zones occurs at different $\eta_{scaled}$ values for the two fluid types: elucidation of this aspect requires further work, as does testing on pitcher fluids.  \newline

It is to be noted that the current theory has not been tested for the pitcher fluids. The critical requirements to perform these experiments on pitcher fluids or any other biological fluids, are as follows: (i) ample amount of fluid should be available (approximately 3-5 mL) for enough number of tests across a wide range of parameters; (ii) measurements of wetting and dispersive properties of the fluid. We invite future studies in this direction, wherein, the experts of pitcher fluids and that of thin-film physics could join forces to further test this theory against pitcher fluids, belonging to different species and locations. \newline

\section{Stability analysis of moving contact line}
\label{sec:mcl}

In previous sections, the condition for formation of droplets or filaments was determined using stability analysis of a young thin-film. We now seek to determine how the spacing between such filaments depends on the contact line velocity. The theory developed in \citep{snoeijer07} is used to derive the scaling-relation for the size and spacing of filaments, which is then compared with our experiments and experimental data reported in the literature. \newline
The dimensionless lubrication equation for a moving contact line of a thin-film of height $h(x,y,t)$ with slip at the fluid-solid interface is given by \citep{oron} 
\begin{subequations}\label{eq:mcl}
\begin{gather}
 \partial_{t} h+ \nabla. (h \ \textbf{U})=0 \label{eq:lubri} \\
 \nabla \kappa + \textbf{e}_{y} + \frac{3(-Ca \ \textbf{e}_{y} - \mathbf{U})}{h(h+3l_{s})}=0 
\end{gather}
\end{subequations}
where $\textbf{U}(x,y,t)=U_{x}\textbf{e}_{x}+U_{y}\textbf{e}_{y}$ is the depth-averaged fluid velocity inside the film, $l_{s}$ is the slip length to ensure Navier-slip along the substrate, $Ca$ is the capillary number given by $\hat{\eta}(\dot{\gamma}) \hat{U}/\hat{\gamma}$, $\hat{U}$ is the contact line velocity, and $\nabla = \textbf{e}_{x}\partial_{x}+\textbf{e}_{y}\partial_{y}$. All the terms are dimensionless. A linear stability analysis of \eqref{eq:mcl} yields
\begin{subequations} \label{eq:mcl_2}
\begin{gather}
  h(x,y,t)= h_{0}(y)+  \epsilon h;_{1}(y)e^{-\sigma t+ iqx} \\
  \kappa(x,y,t)= \kappa_{0}(y)+ \epsilon \kappa_{1}(y)e^{-\sigma t+ iqx}
\end{gather}
\end{subequations}
where $\kappa$ is two times the mean curvature of the interface (nondimensionalised by $1/\bar{h}$). $h_{0}(z)$ and $\kappa_{0}(z)$ denote the shape of the film as a result of balance between capillary forces and gravity; $\hat{\sigma}$ is the growth rate of perturbations; $\hat{q}$ is the wavenumber of the disturbance in the $x$ direction (Fig. \ref{fig:fig3}(a)). The detailed stability analysis of  \eqref{eq:mcl_2} is given in the Appendix of \citep{snoeijer07}, wherein the following analytical result for $\hat{\sigma}$ is derived 
\begin{equation} \label{eq:growth}
    \hat{\sigma} = \frac{|\hat{q}| \hat{\gamma} }{\hat{\eta}} \frac{\tan \theta_{r}}{\cos \theta_{r}}\left(\frac{d \tan \theta_{r}}{d Ca} \right)^{-1}
\end{equation}
where $\theta_{r}$ is the receding contact angle of the thin film, estimated using the Cox-Voinov law \citep{cox},
\begin{equation}\label{eq:cox_voinov}
    \theta_{r}^{3}=\theta_{e}^{3} - 9 \ Ca \ \ln(y/l_{s})
\end{equation}
where $\theta_{e}$ is the equilibrium contact angle, $y$ is the distance from the contact line (Fig. \ref{fig:fig3}(a)) and $l_{s}$ is the slip length. Differentiating w.r.t. $Ca$ gives
\begin{equation}\label{eq:cox}
    3 \theta_{r}^{2} \ \frac{d \theta_{r}}{d Ca}= - 9 \ \ln(y/l_{s}) \frac{d(y/l_{s})}{d Ca}
\end{equation}
such that $\frac{d(y/l_{s})}{d Ca} \propto \hat{T}_{Y}$ because $l_{s}$ is dependent only on molecular factors (self-diffusion coefficient and molecular size) \cite{cox_voi}. Since $\theta_{r}$ is small, $\tan \theta_{r} \approx \theta_{r}$ and \eqref{eq:growth} becomes
\begin{equation}\label{eq:growth_2}
    \hat{\sigma} = - |\hat{q}| \frac{\hat{\gamma}}{\hat{\eta}} \frac{\theta_{r}}{1-\theta_{r}^{2}/2} \left(\frac{3}{\theta_{r}^{2}} \ln\left(\frac{y}{l_{s}}\right) \hat{T}_{Y}^{-1}\right)  \approx -|\hat{q}| \frac{\hat{\gamma}}{\hat{\eta}} \frac{\theta_{r}^{3}}{3 \ln(y/l_{s}) \hat{T}_{Y}}
\end{equation}

\begin{figure}
 \centering
\includegraphics[width=0.9\linewidth]{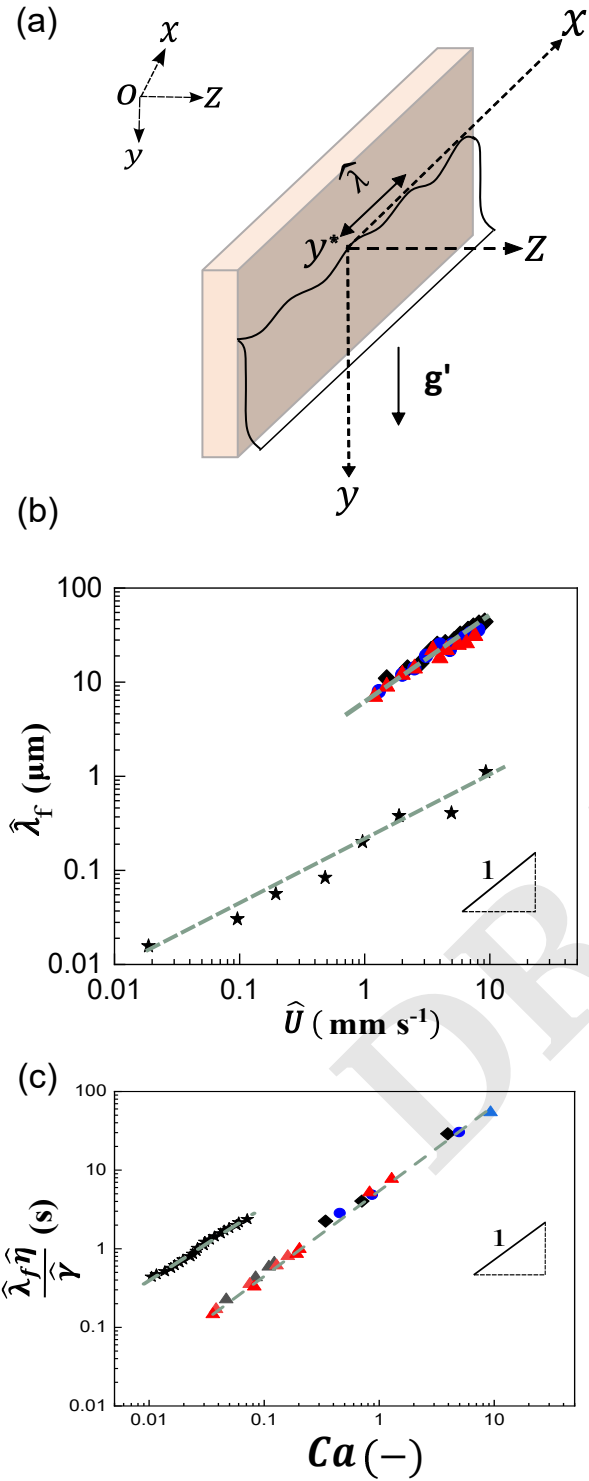}
\caption[Filament spacing: theory and experiments]{Comparison of the \textbf{experimental data from literature and experiments in the current work} on filament spacing, $\hat{\lambda}$\textsubscript{f} with predictions. (a) Schematic of a moving contact line with disturbance wavelength $\hat{\lambda}$ in the $x$ direction. $g'$ is the component of gravity acting along the plane of the substrate; (b) Effect of receding contact line velocity $\hat{U}$ on spacing between filaments $\hat{\lambda}\textsubscript{f}$ for experiments performed by us and reported in the literature \cite{deblais}; (c) Effect of capillary number $Ca$ on $\frac{\hat{\lambda}\textsubscript{f}\hat{\eta}}{\hat{\gamma}}$ for PEO solutions; symbol colour indicates angle of inclination: \textbf{black} for $\alpha=0^{\circ}$; \textcolor{blue}{\textbf{blue}} for $\alpha=30^{\circ}$; \textcolor{red}{\textbf{red}} for $\alpha=60^{\circ}$. Star symbols denote the experimental values reported in the literature \cite{deblais}.} 
\label{fig:fig3}    
\end{figure}

Ultimately, the expression for wavelength $\hat{\lambda}$ of the contact line disturbance, using \eqref{eq:growth_2}, is given by 
\begin{equation}\label{eq:wavelength_1}
    \hat{\lambda} = \frac{2 \pi}{|\hat{q}|} = \frac{2 \pi \hat{\gamma}}{\hat{\sigma} \hat{\eta}} \frac{-\theta_{r}^{3}}{3 \ln (y/l_{s}) \hat{T}_{Y}}  
\end{equation}
Substituting \eqref{eq:cox_voinov} in \eqref{eq:wavelength_1} gives
\begin{equation}
    \hat{\lambda} = \frac{2 \pi \hat{\gamma}}{\hat{\sigma} \hat{\eta} \hat{T}_{Y}} \left[ 3 Ca - \frac{\theta_{e}^{3}}{3 \ln (y/l_{s})} \right]
\end{equation}
with the first term on the RHS given by $\frac{6 \pi \hat{\gamma}}{\hat{\sigma} \hat{T}_{Y} \hat{\eta}} Ca = \frac{6\pi}{\hat{\sigma}\hat{T}_{Y}} \hat{U}$. This predicts that $\hat{\lambda}$ is dependent on the first power of $\hat{U}$ for a given growth rate $\hat{\sigma}$. It should be noted that growth rate in \eqref{eq:growth} is positive, because receding angle $\theta_{r}$ increases with $Ca$ \citep{snoeijer07}. Hence, any small perturbation $\epsilon$ on $h$ will eventually die with a relaxation rate $\hat{\sigma}$, which can be assumed to be constant for a given disturbance $\epsilon$. Interestingly, a similar kind of stability analysis is performed while deriving the growth rate of a Plateau-Rayleigh instability; the major difference being that the wavelength of the most unstable mode is analytically derived in that case, because of imaginary values of growth rate $\hat{\sigma}$. In the present case, we have shown that wavelength $\hat{\lambda}$ is dependent on the first power of $\hat{U}$. \newline

In \S\ref{sec:finitetime}, it was shown that the thin-film, with a thickness lower than the critical film thickness $\hat{h}_{f}$, can break apart, self-similarly, to form filaments. The origin of filament formation is then followed by the motion of contact line and formation of fluctuation-induced crests and troughs of wavelength $\hat{\lambda}$ (Fig. \ref{fig:fig3}(a)). Filaments are experimentally observed to have a regular spacing between them, given by $\hat{\lambda}_{f}$.  \newline

We now compare the above theory with: (i) experimental data collected using PEO solutions in the setup in Figure \ref{fig:fig1}, and (ii) the experimental data reported for aqueous polyacrylamide solutions in \citep{deblais}. For the former, the spacing between filaments $\hat{\lambda}_{f}$ varies with $\hat{U}$ as: $ \hat{\lambda}_{f} \propto \hat{U}^{0.72}$ (Fig. \ref{fig:fig3}(b)) , whereas, for the latter, it varies as $\hat{\lambda}_{f} \propto \hat{U}^{0.80}$ (Fig. \ref{fig:fig3}(c)). In terms of $Ca$, for the former (\textit{i.e.} our experiments), plot of $\hat{\lambda}\textsubscript{f}\hat{\eta}/\hat{\gamma}$ versus $Ca$ shows the following relation: $\hat{\lambda}\textsubscript{f}\hat{\eta}/\hat{\gamma} \propto Ca^{1.08}$ in our experiments and $\hat{\lambda}\textsubscript{f}\hat{\eta}/\hat{\gamma} \propto Ca^{0.945}$ in the literature. There is therefore good agreement between the experiments reported here, those of Deblais \textit{et al.} \cite{deblais}, result obtained here from stability analysis. \newline

\section{Discussion and conclusions}\label{sec:discussion}
The ability to predict regimes of filament/drop formation by a thin-film is likely the result of simplifying the geometry of this problem and focusing on one single question: what drives the formation of these thin-film instabilities? The answer to this question inherently lies in the \textit{finite-time} singularity feature of thin-film PDE which is a 4th order parabolic degenerate PDE. The stability analysis of this equation allows us to find an analytical criterion to predict the conditions which makes the thin-film unstable. Once it happens, the thin-film dynamically evolves to touch the substrate and creates a dry spot region which grows in size and divides the film into multiple filaments. The spacing between filaments is explained by performing stability analysis on the moving contact line, to find out how $\hat{\lambda}_{f}$ depends on $\hat{U}$. \newline

After numerically indicating the finite-time rupture feature of thin-film using sinusoidal perturbation. By applying sinusoidal perturbation to the initial film height in the thin-film equation, the numerical results show that the thin-film tends towards a rupture ($h=0$) in a finite-time. After this, an important question to consider is whether this rupture occurs for any arbitrary perturbation, which is smooth and belongs to $C^{\infty}$ class of function (i.e., can be differentiated infinite times). Pinching-singularities for different thin-film  equations have been proposed previously, but only some have been proven to date. One recent example is the Hele-Shaw case, with $f(h)=h, g(h)=0$ in \eqref{eq:4thordernonlinear}, for which it was numerically shown by \citep{constantin93} that $h \rightarrow 0$ in finite-time, and was proved in 2018 by \citep{constantin18} that the thin-film height $h$ reaches a minimum after a finite-time, which \textit{is} zero. In the similar spirit, the following question remains unsolved:
\begin{itemize}
\item For \eqref{eq:4thordernonlinear} with $f(h)=h^{3}$ and $g(h)=h^{-1}$, prove that for all profiles of initial data, the thin-film height $h$ after a finite-time reaches a minima, \textit{i.e.} zero, giving rise to finite-time singularity.
\end{itemize}

One possible way to prove finite-time singularity for the above problem is to embed this form of \eqref{eq:4thordernonlinear} (hereafter, referred to as the filament equation) on a Turing machine. The techniques on how to do this for Euler equations were reported recently \citep{cardona}. If the same can be done for the filament equation, then it means that for a general initial perturbation, the solution to the equation is computable or, in other words, \textit{Turing-complete}. A Turing-complete solution can be used to program the fluid flow according to the requirement of the problem, as mentioned in \citep{tao2016finite}. The physical interpretation of a Turing-complete flow is that the path of particles are undecidable \footnote{An undecidable problem is a problem for which it is proved to be impossible construct a function/algorithm that always yield a correct `yes' or `no' answer.}  in a sense that nothing \textit{a priori} can be said about their future trajectory; this feature is also known as \textit{universality} of thin-film equation.  Dimensional analysis of the filament equation \citep{witelski2000} shows that the solution behaves in a self-similar way, because of its \textit{scalar-invariant} nature. Combining the \textit{university} feature with the \textit{scalar-invariant} feature, the computable solution of the thin-film equation should evolve in a self-similar fashion for an arbitrary initial datum. The height of the thin-film reduces over time, following the trend shown in Supplementary Figure S4, and then it eventually blows up in a finite-time $t_{f}$ at height $h=0$, for any arbitrary initial datum. The physical manifestation of this singularity is in the origin of the filament formation. Detailed discussion of this point is given in Supplementary Note 2. \newline

Apart from the theoretical-computational treatment as the next step in this work, it is also possible to solve \eqref{eq:4thordernonlinear} using an operator-learner based PDE solver. A physics-informed neural network (PINN) technique was recently employed to find a smooth self-similar solution (scalar invariant) for Boussinesq equations (waves of wavelength $\lambda$ in a shallow water pool of depth $d$; $\lambda \gg d$) \cite{wang2022self}. Whilst the PINN technique can be used to solve \eqref{eq:4thordernonlinear} numerically, the problem of dealing with a singularity requires a resolution-invariant tool. Attention-based operator learners are reported to have this capability, hence any blowup of the solution, if it occurs, will not be dependent on the resolution of the software \cite{cao2021choose}. This is essential if one needs to use a numerically-computed self-similar solution of thin-film PDE, which blows up in a resolution-invariant way, to devise a computer-assisted proof of finite-time singularity formation from an initial datum \cite{gomez2019computer}.

\section{Materials and Methods}
\label{sec:mat_method}
\subsection{Experimental setup}
\label{sec:setup}
The experimental setup consists of a coated glass slide, placed on a stand as shown in Fig. \ref{fig:fig1}(a). A bath consisting of the test liquid is raised vertically upwards to wet the initially dry glass slide, and then brought downwards rapidly to its initial position to generate a thin-film which drains downwards under the action of gravity $\textbf{g}$. Camera $F$ is used to capture the thin-film in the $xy$ plane, and side camera $S$ is used to capture the macroscopic receding contact angle $\theta_{r}$ of the thin-film in the $yz$ plane. Movies of the receding film are recorded using a Ximea xiQ USB3.0 ($F$) camera at 17 fps and a Basler acA1300-200um USB3.0 ($S$) camera at 50 fps. Subsequent image analysis is performed using ImageJ software. All experiments are performed at a lab temperature of $21^{\circ}$C. \newline

Borosilicate glass slides with dimensions $25 \times 75$ mm$^{2}$  and thickness 1 mm were coated with PDMS (polydimethylsiloxane) solution. The PDMS solution was prepared by dissolving Sylgard\textsuperscript{TM} 184 silicone elastomer and curing agent in the weight ratio 10:1. Slides were prepared using the following protocol: cleansing with surface active cleaning agent (Decon\textsuperscript{\textregistered}90 at $2\%$), rinsing with pure water, drying with an air gun, and further cleaning using an air plasma cleaner. PDMS solution was then coated on the glass slide by spin-coating at 3000 rpm for 30 s. 

\subsection{Solution preparation}
\label{sec:material}
PEO solutions (3500 to 11000 ppm) were prepared by adding the required amount of PEO powder (mol. wt. $8 \times 10^{6}$, Acros Organics) to deionised water at room temperature and stirred at 800 rpm for 48 hours to obtain a uniform solution. Glycerol-water mixtures (80-100 $\%$ v/v) were stirred for shorter periods.

\subsection{Material characterisation}
\label{sec:characterisation}
This section describes the characterisation of the liquids employed in the current work. These can be broadly classified as (i) Newtonian, glycerol-water mixtures, (ii) non-Newtonian, aqueous PEO solutions.
\subsubsection{Shear rheology}
\label{sec:shear_rheo_material}
The shear rheology of the liquids was studied using a Kinexus controlled shear rheometer (Malvern Instruments, UK) fitted with a Couette cell. The apparent viscosity,  $\eta_{app}$, was measured using controlled stress ramps. Supplementary Fig. S1(a) shows that the glycerol-water mixtures were Newtonian. The PEO solutions exhibit shear-thinning behaviour, particularly in the experimentally-relevant shear rate range of $0.1-60 \, s^{-1}$. Supplementary Fig. S1(b) shows the fit of the PEO data to the Crosss model, $\frac{\hat{\eta}-\hat{\eta}_{\infty}}{\hat{\eta}_{0}-\hat{\eta}_{\infty}}=  \frac{1}{1+(\lambda_{c} \dot{\gamma})^{m}}$. The model parameters are reported in Table S1 and the values of $\hat{\eta}_{0}$ and $\lambda_{c}$ for different PEO concentrations are plotted in Supplementary Fig. 2. 

\subsubsection{Receding contact angle and surface tension}
\label{sec:ca_st}
The receding angle $\theta_{r}$ of the thin-film was measured using the side camera $S$, as shown in the schematic, Fig. \ref{fig:fig1}(d). The surface tension $\hat{\gamma}$ of the liquids was measured using pendant droplet tensiometry (see \citep{victor}). The values obtained are reported in Supplementary Table S1.

\acknow{We wish to thank Prof. Walter Federle and Dr Victor Kang for their preliminary experiments on filament formation with pitcher plant fluids which inspired this work. The experimental work benefitted from insights provided by Victor Kang, Sagnik Middya, Ratul Das, Sundeep Vema, and Suraj Pavagada on rig development and surface coating techniques. Discussions with Dr A.J.D. Shaikeea, Dr Shuvrangsu Das and a summer school organised by UMass Amherst on Soft Solids and Complex Fluids provided inputs to the theoretical side of this work.}

\subsection*{Author Affiliations}
Department of Chemical Engineering and Biotechnology, University of Cambridge, Philippa Fawcett Drive, Cambridge CB3 0AS, UK
\showacknow{}
\subsection*{Bibliography}
\bibliography{pnas-sample}

\begin{thebibliography}{10}

\bibitem{adlassnig2010deadly}
W Adlassnig, T Lendl, M Peroutka, I Lang, {\em Deadly glue—adhesive traps of
  carnivorous plants}.
\newblock (Springer), (2010).

\bibitem{deblais}
A Deblais, R Harich, A Colin, H Kellay, Taming contact line instability for
  pattern formation.
\newblock {\em\protect\JournalTitle{Nature Communications}} \textbf{7}, 1--7
  (2016).

\bibitem{ytstone}
H Stone, Modern applications of classical ideas in fluid mechanics
  (\url{https://youtu.be/s4Sqml7Kcec?t=3230}) (2021).

\bibitem{witelski2000}
TP Witelski, AJ Bernoff, Dynamics of three-dimensional thin film rupture.
\newblock {\em\protect\JournalTitle{Physica D: Nonlinear Phenomena}}
  \textbf{174}, 155--176 (2000).

\bibitem{witelski2001}
AL Bertozzi, G Grün, TP Witelski, Dewetting films: bifurcations and
  concentrations.
\newblock {\em\protect\JournalTitle{Nonlinearity}} \textbf{14}, 1569 (2001).

\bibitem{witelski2008}
MB Gratton, TP Witelski, Coarsening of unstable thin films subject to gravity.
\newblock {\em\protect\JournalTitle{Phys. Rev. E}} \textbf{77}, 016301 (2008).

\bibitem{witelski2020}
TP Witelski, Nonlinear dynamics of dewetting thin filmss.
\newblock {\em\protect\JournalTitle{AIMS Mathematics}} \textbf{5}, 4229--4259
  (2020).

\bibitem{nasehi2018evolution}
R Nasehi, E Shirani, Evolution of thin liquid film for newtonian and power-law
  non-newtonian fluids.
\newblock {\em\protect\JournalTitle{Scientia Iranica}} \textbf{25}, 266--279
  (2018).

\bibitem{strogatz}
SH Strogatz, {\em Nonlinear dynamics and chaos with student solutions manual:
  With applications to physics, biology, chemistry, and engineering}.
\newblock (CRC press), (2018).

\bibitem{laugesen}
RS Laugesen, MC Pugh, Linear stability of steady states for thin film and
  cahn-hilliard type equations.
\newblock {\em\protect\JournalTitle{Archive for rational mechanics and
  analysis}} \textbf{154}, 3--51 (2000).

\bibitem{stone}
N Xue, HA Stone, Self-similar draining near a vertical edge.
\newblock {\em\protect\JournalTitle{Phys. Rev. Let.}} \textbf{125}, 064502
  (2020).

\bibitem{hocherman}
T Hocherman, P Rosenau, On {KS}-type equations describing the evolution and
  rupture of a liquid interface.
\newblock {\em\protect\JournalTitle{Physica D: Nonlinear Phenomena}}
  \textbf{67}, 113--125 (1993).

\bibitem{bertozzi}
AL Bertozzi, MC Pugh, Long‐wave instabilities and saturation in thin film
  equations.
\newblock {\em\protect\JournalTitle{Comm. on Pure and App. Math.}} \textbf{51},
  625--661 (1998).

\bibitem{limat}
T Podgorski, JM Flesselles, L Limat, Corners, cusps, and pearls in running
  drops.
\newblock {\em\protect\JournalTitle{Phys. Rev. Let.}} \textbf{87}, 036102
  (2001).

\bibitem{snoeijer07}
JH Snoeijer, B Andreotti, G Delon, M Fermigier, Relaxation of a dewetting
  contact line. {P}art 1. {A} full-scale hydrodynamic calculation.
\newblock {\em\protect\JournalTitle{Journal of Fluid Mechanics}} \textbf{579},
  63--83 (2007).

\bibitem{oron}
A Oron, SH Davis, SG Bankoff, Long-scale evolution of thin liquid films.
\newblock {\em\protect\JournalTitle{Reviews of modern physics}} \textbf{69},
  931--976 (1997).

\bibitem{cox}
RG Cox, The dynamics of the spreading of liquids on a solid surface. {P}art 1.
  {V}iscous flow.
\newblock {\em\protect\JournalTitle{Journal of Fluid Mechanics}} \textbf{168},
  169--194 (1986).

\bibitem{cox_voi}
M Do-Quang, Wetting contact angle
  (\url{https://www.mech.kth.se/~luca/Micro/WettingDynamics_Minh.pdf}) (2010).

\bibitem{constantin93}
P Constantin, et~al., Droplet breakup in a model of the {H}ele-{S}haw cell.
\newblock {\em\protect\JournalTitle{Phys. Rev. E}} \textbf{47}, 4169 (1992).

\bibitem{constantin18}
P Constantin, T Elgindi, H Nguyen, V Vicol, On singularity formation in a
  {H}ele-{S}haw model.
\newblock {\em\protect\JournalTitle{Comm. in Math. Phys.}} \textbf{363},
  139--171 (2018).

\bibitem{cardona}
R Cardona, E Miranda, D Peralta-Salas, F Presas, Constructing {T}uring complete
  {E}uler flows in dimension 3.
\newblock {\em\protect\JournalTitle{Proc. Nat. Acad. Sci.}} \textbf{118},
  e2026818118 (2021).

\bibitem{tao2016finite}
T Tao, Finite time blowup for an averaged three-dimensional {N}avier-{S}tokes
  equation.
\newblock {\em\protect\JournalTitle{Journal of the American Mathematical
  Society}} \textbf{29}, 601--674 (2016).

\bibitem{wang2022self}
Y Wang, CY Lai, J G{\'o}mez-Serrano, T Buckmaster, Self-similar blow-up profile
  for the {B}oussinesq equations via a physics-informed neural network.
\newblock {\em\protect\JournalTitle{arXiv preprint arXiv:2201.06780}}
  \textbf{1} (2022).

\bibitem{cao2021choose}
S Cao, Choose a transformer: Fourier or {G}alerkin.
\newblock {\em\protect\JournalTitle{Advances in Neural Information Processing
  Systems}} \textbf{34}, 24924--24940 (2021).

\bibitem{gomez2019computer}
J G{\'o}mez-Serrano, Computer-assisted proofs in {PDE}s: a survey.
\newblock {\em\protect\JournalTitle{SeMA Journal}} \textbf{76}, 459--484
  (2019).

\bibitem{victor}
V Kang, H Isermann, S Sharma, DI Wilson, W Federle, How a sticky fluid
  facilitates prey retention in a carnivorous pitcher plant ({N}epenthes
  {R}afflesiana).
\newblock {\em\protect\JournalTitle{Acta Biomaterialia}} \textbf{27},
  10705--10713 (2021).

\end{thebibliography}

\end{document}